\documentclass{llncs}
\usepackage{graphicx,multirow}

\newcommand{\Din}{\ensuremath{D_\mathsf{in}}}
\newcommand{\Dout}{\ensuremath{D_\mathsf{out}}}
\newcommand{\psum}{\ensuremath{\mathsf{sum}}}
\newcommand{\search}{\ensuremath{\mathsf{search}}}
\newcommand{\rank}{\ensuremath{\mathsf{rank}}}
\newcommand{\select}{\ensuremath{\mathsf{select}}}
\newcommand{\pstart}{\ensuremath{\mathsf{start}}}
\newcommand{\pend}{\ensuremath{\mathsf{end}}}

\begin{document}

\title{On representing the degree sequences of\\sublogarithmic-degree Wheeler graphs}
\author{Travis Gagie}
\institute{}
\maketitle

\begin{abstract}
We show how to store a searchable partial-sums data structure with constant query time for a static sequence $S$ of $n$ positive integers in $o \left( \frac{\log n}{(\log \log n)^2} \right)$, in $n H_k (S) + o (n)$ bits for $k \in o \left( \frac{\log n}{(\log \log n)^2} \right)$.  It follows that if a Wheeler graph on $n$ vertices has maximum degree in $o \left( \frac{\log n}{(\log \log n)^2} \right)$, then we can store its in- and out-degree sequences $\Din$ and $\Dout$ in $n H_k (\Din) + o (n)$ and $n H_k (\Dout) + o (n)$ bits, for $k \in o \left( \frac{\log n}{(\log \log n)^2} \right)$, such that querying them for pattern matching in the graph takes constant time.
\end{abstract}

\section{Introduction}
\label{sec:introduction}

A Wheeler graph~\cite{GMS17} is a directed edge-labelled graph whose vertices can be ordered such that vertices with no in-edges come first; if $u$ has an in-edge labelled $a$ and $v$ has an in-edge labelled $b$ with $a \prec b$ then $u < v$; if edges $(u, v)$ and $(w, x)$ are both labelled $a$ and $u < w$ then $v \leq x$.  Wheeler graphs are interesting because graphs that arise in some important applications are Wheeler --- such as collections of edge-labelled paths and cycles, tries, and de Bruijn graphs --- and if a graph is Wheeler then we can build a small index for it such that, given a pattern, we can quickly tell which vertices can be reached by paths labelled with that pattern.

The index for a Wheeler graph consists of four components:
\begin{enumerate}
\item a data structure supporting $\psum$ queries on the list $\Dout$ of the vertices' out-degrees, with $\Dout.\psum (i)$ returning the $i$th partial sum of the out-degrees (that is, the sum of the out-degrees of the first $i$ vertices in the Wheeler order);
\item a data structure supporting $\rank$ queries on the list $L$ of edge labels sorted by the edges' origins, with $L.\rank_a (i)$ returning the frequency of $a$ among the first $i$ edge labels;
\item a data structure supporting $\psum$ queries on the list $C$ of the edge labels' frequencies, with $C.\psum (a)$ returning the sum of the frequencies of the edge labels lexicographically strictly less than $a$;
\item a data structure supporting $\search$ queries on the list $\Din$ of the vertices' in-degrees, with $\Din.\search (j)$ returning the largest $i$ such that the sum of the in-degrees of the first $i$ vertices in the Wheeler order is at most $j$.
\end{enumerate}

To see how the index works, first notice that, by the definition of a Wheeler graph, the vertices reachable by paths labelled with a pattern $P$ form a single interval in the Wheeler order.  In particular, all the vertices are reachable by paths labelled with the empty string.  Suppose we have already found the endpoints $V_P.\pstart$ and $V_P.\pend$ of the interval $V_P$ in the Wheeler order containing vertices reachable by paths labelled $P$, and we want to find the endpoints $V_{P\,\cdot\,a}.\pstart$ and $V_{P\,\cdot\,a}.\pend$ of the interval $V_{P\,\cdot\,a}$ in the Wheeler order containing vertices reachable by paths labelled $P \cdot a$, where $\cdot$ denotes concatenation.

We use the first data structure to find the endpoints $E_{P,\,\mathrm{out}}.\pstart =\)
\linebreak
\(\Dout.\psum (V_{P\,\cdot\,a}.\pstart) + 1$ and $E_{P,\,\mathrm{out}}.\pend = \Dout.\psum (V_P.\pend)$ of the interval $E_{P,\,\mathrm{out}}$ in $L$ that contains the labels of the out-edges of the vertices in $V_P$.  We then use the second and third data structures to find the endpoints $E_{P\,\cdot\,a}.\pstart = L.\rank_a (E_{P,\,\mathrm{out}}.\pstart - 1) + 1 + C.\psum (a)$ and $E_{P\,\cdot\,a}.\pend = L.\rank_a (E_{P,\,\mathrm{out}}.\pend) + C.\psum(a)$ of the interval $E_{P\,\cdot\,a}$ in the list of edge labels sorted into lexicographic order with ties broken by origin, that contains the copies of $a$ in $E_{P,\,\mathrm{out}}$.  Finally, we use the fourth data structure to find the endpoints $V_{P\,\cdot\,a}.\pstart = \Din.\search (E_{P\,\cdot\,a}.\pstart)$ and $V_{P\,\cdot\,a}.\pend = \Din.\search (E_{P\,\cdot\,a}.\pend)$ of $V_{P\,\cdot\,a}$.

\begin{figure}
\begin{center}
\begin{tabular}{c@{\hspace{6ex}}c}
\begin{tabular}{c}
\includegraphics[width=.5\textwidth]{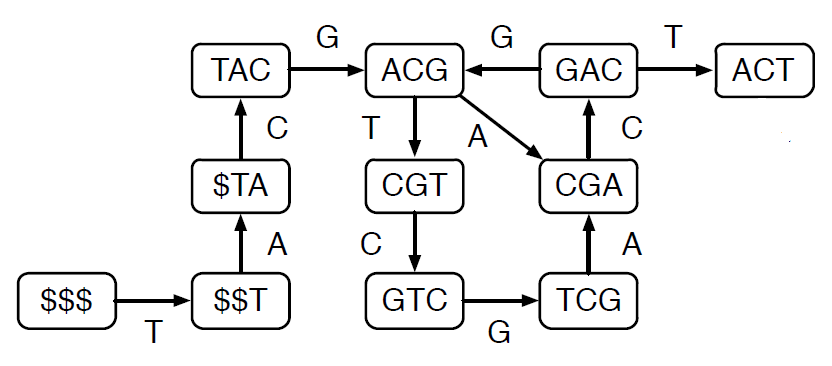} \\[6ex]
\begin{tabular}{c@{\hspace{8ex}}c}
\begin{tabular}{c|c}
a & $C.\psum (a)$ \\
\hline & \\[-1ex]
{\sf A} & 0 \\
{\sf C} & 3 \\
{\sf G} & 6 \\
{\sf T} & 9 \\
\end{tabular}
&
\raisebox{-7ex}{\includegraphics[width=.25\textwidth]{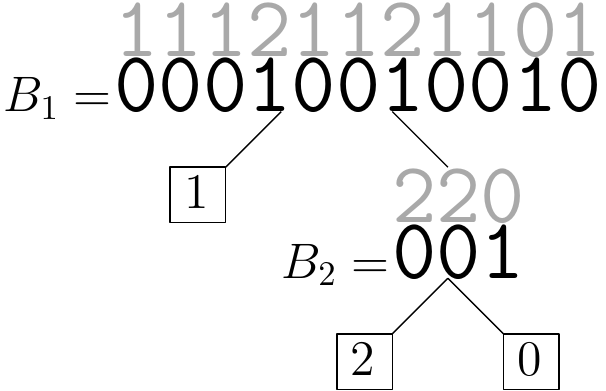}}
\end{tabular}
\end{tabular}
&
\begin{tabular}{rc|ccc}
\multicolumn{1}{c}{Wheeler} \\
\multicolumn{1}{c}{rank}
    & vertex             & $\Dout$ & $L$ & $\Din$ \\
\hline & \\[-1ex]
1)  & \sf \,\$\,\$\,\$\, & 1       & T   & 0 \\[.5ex]
2)  & \sf CGA            & 1       & C   & 2 \\[.5ex]
3)  & \sf \,\$\,TA     	 & 1       & C   & 1 \\[.5ex]
\multirow{2}{*}{4)} & \multirow{2}{*}{\sf GAC} & \multirow{2}{*}{2} & G & \multirow{2}{*}{1} \\
    &                    &         & T   &   \\[.5ex]
5)  & \sf TAC            & 1       & G   & 1 \\[.5ex]
6)  & \sf GTC            & 1       & G   & 1 \\[.5ex]
\multirow{2}{*}{7)} & \multirow{2}{*}{\sf ACG} & \multirow{2}{*}{2} & G & \multirow{2}{*}{2} \\
    &                    &         & T   &   \\[.5ex]
8)  & \sf TCG            & 1       & A   & 1 \\[.5ex]
9)  & \sf \,\$\,\$\,T    & 1       & A   & 1 \\[.5ex]
10) & \sf ACT            & 0       &     & 1 \\[.5ex]
11) & \sf CGT            & 1       & C   & 1
\end{tabular}
\end{tabular}

\bigskip

\caption{A Wheeler graph {\bf (upper left;~\cite{BBGPS15})}; a table with $\Dout$, $L$ and $\Din$ {\bf (right)}; $C$ {\bf (lower left)}; and a degenerate wavelet tree supporting $\psum$ on $\Dout$ {\bf (lower center)}.}
\label{fig:example}
\end{center}
\end{figure}

This works because, again by the definition of a Wheeler graph, the list of edge labels sorted into lexicographic order with ties broken by origin, is also sorted by the ranks in the Wheeler order of the edges' destinations.  For the sake of brevity, however, we refer the reader to Gagie et al.'s~\cite{GMS17} original paper on Wheeler graphs for a full proof of correctness, and offer here only the example in Figure~\ref{fig:example} (modified from~\cite{BBGPS15}), which shows the BOSS~\cite{BOSS12} representation of a de Bruijn graph (with the out-edge leaving vertex {\sf ACT} and labelled {\sf \$} deleted).

Suppose we have already found the endpoints $V_{\sf C}.\pstart = 4$ and $V_{\sf C}.\pend = 6$ of the interval $V_{\sf C}$ of vertices reachable by paths labelled {\sf C}, and we want to find the endpoints $V_{\sf CG}.\pstart = 7$ and $V_{\sf CG}.\pend = 8$ of the interval $V_{\sf CG}$ of vertices reachable by paths labelled {\sf CG}.  We compute
\begin{eqnarray*}
\Dout.\psum (4 - 1) + 1 & = & 4\\
\Dout.\psum (6) & = & 7\\[2ex]
L.\rank_{\sf G} (4 - 1) + 1 + C.\psum ({\sf G}) & = & 7\\
L.\rank_{\sf G} (7) + C.\psum ({\sf G}) & = & 9\\[2ex]
\Din.\search (7) & = & 7\\
\Din.\search (9) & = & 8
\end{eqnarray*}
and correctly conclude $V_{\sf CG} = [7, 8]$ (containing vertices {\sf ACG} and {\sf TCG}).

There have been many papers on how to represent $L$ compactly while supporting fast $\rank$ queries on it, and representing $C$ compactly while supporting fast $\psum$ queries on it is trivial unless the alphabet of edge labels is unusually large, so in this paper we focus on how to represent $\Dout$ and $\Din$ compactly while supporting fast $\psum$ and $\search$ queries on them.  Specifically, we describe the first searchable partial-sums data structure for a static sequence $S$ of sublogarithmic positive integers, with constant query time and space bounded in terms of the $k$th-order empirical entropy $H_k (S)$ of $S$:

\begin{theorem}
\label{thm:sps}
Let $S [1..n]$ be a static sequence of positive integers.  If $\max_i \{S [i]\}, k \in o \left( \frac{\log n}{(\log \log n)^2} \right)$ then we can store $S$ in $n H_k (S) + o (n)$ bits and support $\psum$ and $\search$ queries on it in constant time.
\end{theorem}

Theorem~\ref{thm:sps} may be of independent interest and it is easy to apply to support $\psum$ queries on $\Dout$ and $\search$ queries on $\Din$.  To see how we can apply it to $\Dout$, notice that if $\Dout'$ is the sequence obtained from $\Dout$ by incrementing each out-degree, then $\Dout'$ contains only positive integers, $|\Dout'| H_k (\Dout') = |\Dout| H_k (\Dout)$ and $\Dout.\psum (i) = \Dout'.\psum (i) - i$.  To see how we can apply it to $\Din$, notice that all the 0s in $\Din$ are at the beginning (by the definition of a Wheeler graph), so if $\Din'$ is the sequence obtained from $\Din$ by deleting its leading 0s, then $\Din'$ contains only positive integers, $|\Din'| H_k (\Din') \leq |\Din| H_k (\Din)$ and $\Din.\search (j) = \Din'.\search (j) + |\Din| - |\Din'|$.  This gives us our main result:

\begin{theorem}
\label{thm:main}
Let $G$ be a Wheeler graph on $n$ vertices with maximum degree $\Delta$.  If $\Delta, k \in o \left( \frac{\log n}{(\log \log n)^2} \right)$ then we can store $G$'s out-degree sequence $\Dout$ in $n H_k (\Dout) + o (n)$ bits such that it supports $\psum$ queries in constant time, and store $G$'s in-degree sequence $\Din$ in $n H_k (\Din) + o (n)$ bits such that it supports $\search$ queries in constant time.
\end{theorem}

\section{Intuition}
\label{sec:intuition}

The standard approach, proposed by M\"akinen and Navarro~\cite{MN07}, to storing a compact searchable partial-sums data structure for a static sequence $S [1..n]$ of positive integers that sum to $u$, is as a bitvector $B$ in which there are $S [1] - 1$ copies of 0 before the first 1 and, for $i > 1$, there are $S [i] - 1$ copies of 0 between the $(i - 1)$st and $i$th copies of 1.  This takes $n \lg \frac{u}{n} + o (u)$ bits and supports $S.\psum (i) = B.\select_1 (i)$ and $S.\search (j) = B.\rank_1 (j)$ in constant time.  If we use it to store the in- and out-degrees in a BOSS representation of a de Bruijn graph then we use about $\lg \sigma + 2$ bits per edge.

There are many other searchable partial-sums data structures (see, e.g.,~\cite{BGS22,PV21} and references therein) but, as far as we know, only a very recent one by Arroyuelo and Raman~\cite{AR22} achieves a space bound in terms of the empirical entropy of $S$ and still answers queries in constant time.  It takes $n H_0 (S) + O \left( \frac{u (\log \log u)^2}{\log u} \right)$ bits so, if $\max_i \{S[i]\} \in o \left( \frac{\log n}{(\log \log n)^2} \right)$, then $u \in o \left( \frac{n \log n}{(\log \log n)^2} \right)$ and it takes $n H_0 (S) + o (n)$ bits.  If we apply this instead of Theorem~\ref{thm:sps} then we obtain a slightly weaker form of Theorem~\ref{thm:main}, in which $H_k$ is replaced by $H_0$.

To prove Theorem~\ref{thm:sps}, our starting point is Ferragina and Venturini's~\cite{FV07} well-known result about storing a static string in $n H_k$-compressed space while supporting fast random access to it:

\begin{theorem}[Ferragina and Venturini]
\label{thm:FV07}
We can store $S$ as a string of $n$ characters from an alphabet of size $\sigma$ in
\[n H_k (S) + O \left( \frac{n \log \sigma}{\log n} (k \log \sigma + \log \log n) \right)\]
bits for $k \in o \left( \frac{\log n}{\log \sigma} \right)$ such that we can extract any substring of $S$ of length $\ell$ in $O \left( 1 + \frac{\ell \log \sigma}{\log n} \right)$ time.
\end{theorem}

Assuming $S$ consists of positive integers with $\max_i \{S [i]\} \in o \left( \frac{\log n}{(\log \log n)^2} \right)$, we have $\sigma \in o \left( \frac{\log n}{(\log \log n)^2} \right)$ and the space bound in Theorem~\ref{thm:FV07} is $n H_k (S) + o (n)$ bits for $k \in o \left( \frac{\log n}{(\log \log n)^2} \right)$.  Notice the extraction time is constant for $\ell \in O \left( \frac{\log n}{\log \sigma} \right)$.

In order to support $\psum$ and $\search$ on $S$ in constant time, we augment Ferragina and Venturini's representation of $S$ with sublinear data structures similar to those Raman, Raman and Rao~\cite{RRR02} used to support {\sf rank} and {\sf select} on their succinct bitvectors.  Since these augmentations are fairly standard, we omit the details of the how we support $\psum$ and leave the details of how we support $\search$ to the next section.

\begin{lemma}
\label{lem:sum}
We can add $o (n)$ bits to Ferragina and Venturini's representation of $S$ and support $\psum$ in constant time.
\end{lemma}

\begin{lemma}
\label{lem:search}
We can add $o (n)$ bits to Ferragina and Venturini's representation of $S$ and support $\search$ in constant time.
\end{lemma}

Combining Theorem~\ref{thm:FV07} and Lemmas~\ref{lem:sum} and~\ref{lem:search}, we immediately obtain Theorem~\ref{thm:sps}.  We note that we need $\sigma \in o \left( \frac{\log n}{(\log \log n)^2} \right)$ only to prove Lemma~\ref{lem:search}.  In the full version of this paper we will show how we can store $S$ in $n H_k (S) + o (n)$ bits of space and support $\psum$ queries on it in constant time even when $\sigma$ is polylogarithmic in $n$, for example --- which could be of interest when storing Wheeler graphs with large maximum out-degree but small maximum in-degree, such as some tries.

\section{Proof of Lemma~\ref{lem:search}}
\label{sec:proof}

\begin{proof}
We first store $\search (c \sigma \lg^2 n)$ for each multiple $c \sigma \lg^2 n$ of $\sigma \lg^2 n$.  Since $\psum (n) \leq \sigma n$, this takes a total of
\[O \left( \frac{\sigma n}{\sigma \lg^2 n} \cdot \lg n \right) \subset o (n)\]
bits.  We then store the difference
\[\search \left( c \cdot \frac{\lg n}{2 \lg \sigma} \right) -
\search \left( \sigma \lg^2 (n) \cdot \left\lfloor \frac{c \cdot \frac{\lg n}{2 \lg \sigma}}{\sigma \lg^2 n} \right\rfloor \right)\]
for each multiple $c \cdot \frac{\lg n}{2 \lg \sigma}$ of $\frac{\lg n}{2 \lg \sigma}$ and the preceding multiple $\sigma \lg^2 (n) \cdot \left\lfloor \frac{c \cdot \frac{\lg n}{2 \lg \sigma}}{\sigma \lg^2 n} \right\rfloor$ of $\sigma \lg^2 n$.  Since each of these differences is at most $\sigma \lg^2 n$, this takes a total of
\[O \left( \frac{\sigma n \log \sigma}{\log n} \cdot \log (\sigma \log^2 n) \right)
\subset o (n)\]
bits.  Finally, we store a universal table that, for each possible $\frac{\lg n}{2}$-bit encoding of a substring of $S$ consisting of $\frac{\lg n}{2 \lg \sigma}$ integers (each represented by $\lg \sigma$ bits) and each value $q$ between 1 and the maximum possible sum $\sigma \cdot \frac{\lg n}{2 \lg \sigma}$ of such a substring, tells us how many of that substring's integers we can sum before exceeding $q$.  This takes
\[2^{\frac{\lg n}{2} + \lg \left(\sigma \cdot \frac{\lg n}{2 \lg \sigma} \right)} \lg \left( \frac{\lg n}{2 \lg \sigma} \right)
\in o (n)\]
bits.

To evaluate $\search (j)$ in constant time, we first look up $\search \left( \sigma \lg^2 (n) \cdot \left\lfloor \frac{j}{\sigma \lg^2 n} \right\rfloor \right)$ and
\[\search \left( \frac{\lg n}{2 \lg \sigma} \cdot \left\lfloor \frac{j}{\frac{\lg n}{2 \lg \sigma}} \right\rfloor \right) -
\search \left( \sigma \lg^2 (n) \cdot \left\lfloor \frac{j}{\sigma \lg^2 n} \right\rfloor \right)\,,\]
which tells us  $\search \left( \frac{\lg n}{2 \lg \sigma} \cdot \left\lfloor \frac{j}{\frac{\lg n}{2 \lg \sigma}} \right\rfloor \right)$.  Since
\[j - \frac{\lg n}{2 \lg \sigma} \cdot \left\lfloor \frac{j}{\frac{\lg n}{2 \lg \sigma}} \right\rfloor < \frac{\lg n}{2 \lg \sigma}\]
and the integers in $S$ are positive,
\[\search (j) - \search \left( \frac{\lg n}{2 \lg \sigma} \cdot \left\lfloor \frac{j}{\frac{\lg n}{2 \lg \sigma}} \right\rfloor \right) < \frac{\lg n}{2 \lg \sigma}\,.\]
It follows that we can find $\search (j)$ by extracting the substring of $\frac{\lg n}{2 \lg \sigma} \in O \left( \frac{\log n}{\log \sigma} \right)$ characters starting at $S \left[ \search \left( \frac{\lg n}{2 \lg \sigma} \cdot \left\lfloor \frac{j}{\frac{\lg n}{2 \lg \sigma}} \right\rfloor \right) \right]$ and using the universal table to learn how many of that substring's integers we can sum before exceeding
\[j - \psum \left( \search \left( \frac{\lg n}{2 \lg \sigma} \cdot \left\lfloor \frac{j}{\frac{\lg n}{2 \lg \sigma}} \right\rfloor \right) - 1 \right)\,.\]
\end{proof}

\section{Postscript}
\label{sec:postscript}

We have not implemented Theorems~\ref{thm:sps} or~\ref{thm:main} because there are other approaches that perform poorly in the worst case but are likely unbeatable in practice.  If we store $S$ as a degenerate wavelet tree, then we can implement an $S.\psum$ query with $\sigma$ $\rank$ queries on the wavelet trees bitvectors, together with $\sigma$ multiplications and additions: for example, to find $\Dout.\psum (8)$ for the sequence $\Dout = 1, 1, 1, 2, 1, 1, 2, 1, 1, 0, 1$ with the degenerate wavelet tree shown in Figure~\ref{fig:example}, we compute
\[1 \cdot B_1.\rank_0 (8) + 2 \cdot B_2.\rank_0 (8 - B_1.\rank_0 (8))
 = 1 \cdot 6 + 2 \cdot 2 = 10\,.\]
In practice $\sigma$ is usually a small constant --- often 4 --- and if the bitvectors in the wavelet tree are entropy-compressed, then it takes $n H_0 (S) + o (n \log \sigma)$ bits.  If we store a minimal monotone perfect hash function~\cite{BBPV11} mapping each value $S.\psum (i)$ to $i$, together with a small sample of those pairs, then we should also be able to support $S.\search$ queries by computing a few hash values and $S.\psum$ queries, quickly and in small space in practice.  We leave the details for the full version of this paper.

\subsection*{Acknowledgments}

Many thanks to Jarno Alanko for bringing the topic of this paper to our attention, to Rossano Venturini for pointing out Arroyuelo and Raman's result, and to Meng He, Gonzalo Navarro and Srinivasa Rao Satti for helpful discussions.

\pagebreak

\end{document}